\documentstyle[12pt]{article}
\normalbaselines
\catcode `\@=11
\@addtoreset{equation}{section}

\def\section{\@startsection {section}{1}{\z@}{-3.5ex plus -1ex minus
    -.2ex}{2.3ex plus .2ex}{\normalsize\bf}}
\def\subsection{\@startsection{subsection}{2}{\z@}{-3.25ex plus -1ex minus
 -.2ex}{1.5ex plus .2ex}{\normalsize\bf}}
\def\thebibliography#1{\section*{References\markboth
  {REFERENCES}{REFERENCES}}\list
  {[\arabic{enumi}]}{\settowidth\labelwidth{[#1]}\leftmargin\labelwidth
  \advance\leftmargin\labelsep
  \usecounter{enumi}}
  \def\newblock{\hskip .11em plus .33em minus -.07em}
  \sloppy
  \sfcode`\.=1000\relax}

\catcode `\@=12
\begin{document}
\vspace*{2.5cm}
\noindent
{\bf POLYNOMIAL LIE ALGEBRAS AND ASSOCIATED PSEUDOGROUP
STRUCTURES IN COMPOSITE QUANTUM MODELS}\vspace{1.3cm}\\
\noindent
\hspace*{1in}
\begin{minipage}{13cm}
Valery P. Karassiov \vspace{0.3cm}\\
Lebedev Physical Institute,\\
Leninsky prospect 53, 117924 Moscow,\\
Russia
\end{minipage}

\vspace*{0.5cm}

\begin{abstract}
\noindent
Polynomial Lie (super)algebras $g_{pd}$ are introduced via $G_{i}$-invariant
polynomial Jordan maps in  quantum composite models with Hamiltonians $H$
having invariance groups $G_{i}$. Algebras $g_{pd}$ have polynomial structure
functions in commutation relations, are related to pseudogroup structures
$\exp V,\, V\in g_{pd}$  and describe dynamic symmetry of models under study.
Physical applications of algebras $g_{pd}$ in quantum optics and in composite
field theories are briefly discussed.

\end{abstract}
\section{\hspace{-4mm}.\hspace{2mm}INTRODUCTION}
\hspace*{0.8cm}The symmetry methods are fruitfully exploiting in quantum
physics from the time of its origin (see, e.g., [1-6] and references
therein). In particular, they provide an elegant treatment of physical tasks
using a powerful formalism of Lie groups and algebras, especially generalized
coherent states (GCS) and related technigues [2,4-6], when Hamiltonians $H$
of systems under study are linear forms
\begin{equation}
H=\sum_{\alpha =1}^d \lambda_{\alpha} F_{\alpha}+C,\qquad [F_{\alpha},
F_{\beta}]_- \equiv F_{\alpha} F_{\beta} - F_{\beta} F_{\alpha} =
\sum_{\gamma =1}^d c_{\alpha \beta}^{\gamma}F_{\gamma}\equiv
\psi_{\alpha \beta}^1(\{F_{\gamma}\})
\end{equation}
in generators $F_{\alpha}$ of $d (<\infty)$-dimensional Lie algebras $g^D$ of
dynamic symmetry ($\lambda_{\alpha}$ are $c$-number coefficients and
$[F_{\alpha},C]=0$). But for last years nonlinear models in many branches of
quantum physics have
called for different extensions of usual Lie algebras in Eq. (1.1) by means
of 1) admitting infinite dimensions $d$ [7,8], 2) involving two types
commutators ($[,]_{\mp}$) defining Lie superalgebras [7] and 3) using
nonlinear structure functions $\psi_{\alpha \beta}~(\{F_{\alpha}\})$ defining
nonlinear or deformed Lie (super)algebras [7-11]. However, at present, these
new Lie-algebraic structures, enabling to display important structure
features of models with such generalized Hamiltonians (1.1), do not yield
universal techniques for solving physical tasks due to the absence of simple
("finite") "disentangling" (Zassenhaus) and "multiplication" (Baker-Campbell-
Hausdorff) formulas for their exponentials [5,7,10] which determine a high
efficiency of group-theoretical methods [2,4]. Therefore, keeping in mind
relevant extensions of these methods for new algebras, it is of importance
to examine possibilities of adequate modifications of group-theoretical and
Lie-algebraic techniques for single classes of models.

In the present paper, summarizing and developing results of the papers [5,
12,13], we discuss these problems for polynomial Lie (super)algebras $g_{pd}$
which describe dynamic symmetries of composite many-body models with
Hamiltonians $H$ having invariance groups $G_{i}(H)$ and are obtained via
generalized $G_{i}$-invariant Jordan maps [5].

\section{\hspace{-4mm}.\hspace{2mm}POLYNOMIAL $G_{i}$-INVARIANT
JORDAN MAPS AND POLYNOMIAL LIE (SUPER)ALGEBRAS IN MANY-BODY PHYSICS}

\hspace*{0.8cm}
As is known, in composite models of many-body physics whose Hamiltonians
$H$ and quantum state spaces $L(H)$ are given in terms of boson-fermion
operators $a_i, a_i^+, b_j, b_j^+ ([a_i, a_j^+]_-=\delta_{ij}, [b_j, b_j^+]_+
= \delta_{ij})$, Lie-algebraic methods are introduced in a natural way via
different boson-fermion maps [2-5,14]. Specifically, the Jordan map [5,14]:
$(a_i, a_i^+, b_j, b_j^+)\mapsto F_{\alpha}$ giving generators $F_{\alpha}$
by quadratic forms in $a_i, a_i^+, b_j, b_j^+$ reduces quadratic (in field
operators) Hamiltonians $H_0(a_i, a_i^+, b_j, b_j^+)$ to the form (1.1) [2-5].

This map, introducing collective dynamic variables $F_{\alpha}\in g_0^D$,
is particulariy fruitful when $H_0$ have (both continious and discrete)
invariance groups $G_{i}(H_0)$:
\begin{equation}
[H_0, G_{i}(H_0)]_-=0 \quad\Longrightarrow \quad [g_0^D, G_{i}(H_0)]_-=0
\end{equation}
and field operators are transformed with respect to certain (as a rule,
fundamental) irreducible representations (IRs) of the groups $G_{i}(H_0)$
[5,15]. Then $F_{\alpha}\in g_0^D$ are quadratic vector $G_{i}(H_0)$-
invariants; besides, by virtue of the construction and Eq. (2.1), invariant
(Casimir and class) operators $C_k(G_{i})$ and $C_k(g_0^D)$ determining IRs
of $G_{i}$ and $g_0^D$ are functionally connected and their eigenvalues on
spaces $L(H)$ are specified by certain common sets $[l_i]\equiv [l_0,l_1,
\dots]$ of invariant quantum numbers $l_i$ labeling extremal (lowest) vectors
$|[l_i]\rangle$ of both $G_{i}$ and $g_0^D$ IRs. All that, in turn, entails
spectral decompositions
\begin{equation}
 L(H) =\sum_{[l_i]}\;\sigma([l_i])\; L([l_i])
\end{equation}
of spaces $L(H)$ in direct sums of the subspaces $L([l_i])$ which are
invariant with respect to joint actions of algebras $g_0^D$ and groups
$G_{i}$ being carrier-spaces of so-called factor-representations (isotypic
components) [2] of both algebraic structures $G_{i}$ and $g_0^D$.

In the case of suitable groups $G_{i}$ decompositions (2.2) have the simple
spectra ($\sigma([l_i])=1$), and, then, pairs $(G_{i}, g_0^D)$ say to act
complementarily [15] on $L(H)$ and to form the Weyl-Howe dual pairs since
such pairs were first considered in quantum mechanics by H. Weyl in the
analysis of interrelations between unitary and permutation symmetries of
$N$-electron systems [1], and their explicit mathematical characterization
was given by R. Howe [16]. A physical importance of dual pairs is due to that
they describe completely both invariance and dynamic symmetries of models
under study (see, e.g., [15,5,12] and references therein).

The constructions above are generalized in a natural manner when extending
quadratic Hamiltonians $H_0$ by $G_{i}$-invariant polynomials $H_I(a_i, a_i^+,
b_j, b_j^+)$ of higher degrees which describe essentially nonlinear
interactions [5,17]. Then generalized dual pairs $(G_{i}, g_{pd})$ of
invariance groups $G_{i}$ and (describing dynamic symmetry) Lie-like
(super)algebras $g^D=g_{pd}$ are obtained via $G_{i}$-invariant generalized
Jordan maps [5]
\begin{equation}
(a_i, a_i^+, b_j, b_j^+) \longmapsto (F_{\alpha}, V_{\lambda}, V^+_{\lambda})
\end{equation}
expanding the sets $\{F_{\alpha}\in g_0^D\}$ by some additional generators
$V_{\lambda}, V^+_{\lambda}$ which are simultaneously elementary vector
$G_{i}$-invariants and components of two mutually contragradient $g_0^D$-
irreducible tensor operators $V, V^+$ given by homogeneous polynomials in
$a_i, a_i^+, b_j, b_j^+$. (Note that in practice Hamiltonians $H_0, H_I$
may contain, besides $a_i, a_i^+, b_j, b_j^+$, other $g_0^D$-covariant
operators, e.g., the Pauli matrices $\sigma_{\alpha}(i)$ etc. that leads to
appropriate modifications of Eq.(2.3) [5,12].) Then, by virtue of the vector
invariant theory [1], the sets $\{F_{\alpha}, V_{\lambda}\}$ form
finite-dimensional integrity bases [1] of associative algebras
${\cal A}_{G_{i}}$ of $G_{i}$-invariants embedded in enveloping algebras
${\cal U}(w(m))$ of the Weyl-Heisenberg (super)algebras $w(m)$ with
generators $a_i, a_i^+, b_j, b_j^+$. Furthermore, endowing sets
$\{F_{\alpha}, V_{\lambda}\}$ by commutators $[,]_{\pm}$, one gets
(via a specific extension of the Ado's theorem [2]) {\bf finite-dimensional}
Lie-like (super)algebras $g_{pd}=g^D$ of dynamic symmetry which, however,
have polynomial structure functions $\psi^p_{\alpha \beta}(\{F_{\alpha}\})
(p=deg(\psi))$ for commutators $[V_{\alpha},V_{\beta}]_{\pm}, [V_{\alpha},
V^+_{\beta}]_{\pm}$ (the subscripts "$\pm$" are determined by the "fermion
contents" of operators $V_{\alpha}$) and may be named as polynomial Lie
(super)algebras. Emphasize an importance of the $G_{i}$-invariance of
polynomials $H_I(\dots)$ because, in general, it is impossible to get
finite-dimensional algebras if cancelling this condition [5,12].

Algebras $g_{pd}$ are extensions of Lie algebras $g_0^D$ and have the coset
structure [10,11]:
\begin{equation}
g_{pd} = h + v, \quad h=g_0^D,\quad v=Span\{V_{\alpha}, V^+_{\alpha}\},
\qquad [h, v]\subseteq v,\quad [v, v]\subset {\cal U}(h)
\end{equation}
that enables us to construct IRs of $g_{pd}$ starting from $h$-modules [5]
(${\cal U}(h)$ are enveloping algebras of $h$, and hereon we omit the
subscript "$\pm$" in $[,]_{\pm}$). For example, extensions via (2.3) of the
unitary algebras $u(m)$ by their ($C_n$-invariant) symmetric and
($SU(n)$-invariant) skew-symmetric tensor operators give two classes  of
polynomial oscillator (super)algebras (see Section 4) wheras such extensions
of the symplectic algebras $sp(2m,R)$ by $SO(n)$-invariant skew-symmetric
tensors yield polynomial deformations of the Lie algebra $u(m,m)$ [5].
Without dwelling on other general properties of algebras $g_{pd}$ we only
note that they are close in their structure with $W_n$-algebras [10], can be
enlarged (via repeated commutators) to certain graded infinite-dimensional
Lie (super)algebras $\hat{g}_{pd}=\sum_{r=0}^{\infty}g_r,\,[g_r,g_s]\subset
g_{r+s},\,g_{r\geq 0}={\cal U}(h)(v)^r$, and their exponentials
$\exp (g_{pd})$ generate (non-analytical) pseudogroup structures having, in
general, no finite disentangling formulas [5]. And now we consider
applications of concrete algebras $g_{pd}$.

\section{\hspace{-4mm}.\hspace{2mm}POLYNOMIAL LIE ALGEBRAS $sl_{pd}(2)$ IN
QUANTUM OPTICS}

\hspace*{0.8cm}Simplest examples of polynomial Lie algebras are given by
algebras $sl_{pd}(2)=Span(V_0,V_{\pm})$, obtained via extending the unitary
algebra $u(1)=Span(V_0)$ by generators $V_{\pm}$ and satisfying the
commutation relations (CRs)
%%%%%%%%%
\begin{eqnarray}
[V_0, V_{\pm}]= \pm V_{\pm}, \; [V_-, V_+] = \psi^{p-1} (V_0)\equiv
\Psi^p (V_0+1) -\Psi^p (V_0),\; [\Psi^p(R_0), V_{\alpha}]=0
\end{eqnarray}
where $\Psi^p (\dots)$ is the polynomial of degree $p$ in the variable $V_0$,
$\Psi^p(R_0)=\Psi^p(V_0)-V_+V_-$ is the $sl_{pd}(2)$ Casimir operator (with
$R_0$ being the "lowest weight operator") and hereon we omit the identity
operator symbol $I$ in expressions like $\Psi^p (V_0 + \alpha I)$. As is seen
from Eq. (3.1), algebras $sl_{pd}(2)$ are reduced to the Lie algebra $sl(2)$
when $p=2$ and may be considered as its specific deformations; they are also
obtained from certain $q$-deformed algebras by means of the Wigner-
In$\ddot{o}$nu contraction when $q\rightarrow 1$ [6].

Algebras $sl_{pd}(2)$ arise via the map (2.3) in nonlinear models of quantum
optics where coset generators $ V_{\pm}$ are interpreted as creation/
destruction operators of specific coherent structures (clusters) [12].
For example, Hamiltonians
%=====
\begin{eqnarray}
H= H_0+H_I= \sum _{i=0}^1 \omega_i a_i^+ a_i + g (a^+_1)^n (a_0)^m +
g^* (a^+_0)^m (a_1)^n, \;m\leq n,
\end{eqnarray}
describing multiphoton processes of scattering on multimode Fock spaces $L_F
\equiv L(H)=Span\{|\{n_i\}\rangle=\prod_i [n_i!]^{-1/2}(a_i^+)^{n_i}|0>\}$
($g$ are coupling constants, $\omega_i$ are field modes frequencies and
$\hbar=1$), have invariance groups $G_i(H)= C_{n}\otimes C_{m}\otimes
\exp (i\lambda R_1)$ where $C_{n}=\{c_{kn}=\exp (i2\pi k/n): a_i^+\rightarrow
c_{kn} a_i^+\}, R_1 = (ma^{+}_1 a_1+na_0^+a_0)/(m+n)$.

Then the map (2.3) given as follows
%%%%
\begin{eqnarray}
V_0= (a^+_1 a_1-a_0^+a_0)/(m+n),\; V_+ = (a_1^+ )^n(a_0)^m,
\quad V_- =(V_+)^+
\end{eqnarray}
reduces Eq. (3.2) to the form
%%%%%%%%%%%%%%
\begin{eqnarray}
H=aV_0 + g V_+ + g^* V_- +C,\quad [V_{\alpha}, C]=0,\; a= n\omega_1-m\omega_0,
\, C= R_1(\omega_1+\omega_0)
\end{eqnarray}
and determines the generalized dual pairs ($G_i(H), sl_{pd}(2))$. The
structure polynomials $\Psi^p (V_0)$ are determined with the help of Eqs.
(3.3), the characteristic relation
\begin{equation}
(V_+V_- -\Psi^p (V_0))|_{L(H)}=0
\end{equation}
and defining relations for $a(i), a^+(i)$ [12]:
%========
\begin{equation}
\Psi^p(V_{0})=(n V_0 + R_1)^{(n)} (R_1-mV_0 + m)^{(m)},\; p=m+n,
\, A^{(b)}=A(A-1)...(A-b+1)
\end{equation}
%%%%%
(An extra dependence of $\Psi^p (V_0)$ on $R_1$ reflects functional
interrelations between invariant operators of $G_{i}(H)$ and $sl_{pd}(2)$.)
The subspaces $L([l_i])\equiv Span\{|[l_i]; v\rangle=\propto V_+^v
|[l_i]\rangle,\, V_0|[l_i];v\rangle=(l_0 + v)|[l_i]; v\rangle, \,R_i|[l_i];
v\rangle = l_i |[l_i];v\rangle, i=0,1, \,V_-|[l_i]\rangle=0\}$ in Eq. (2.2)
are generated by the lowest vectors $|[l_i]>=[s!\kappa !]^{-1/2}(a_0^+)^s
(a_1^+)^{\kappa}|0>,\;\kappa= 0,1,\dots,n-1,\; s=0,1,...$ where $l_0=
(\kappa-s)/(m+n), l_1=(m\kappa+ns)/(m+n)$; $\kappa$ specifies IRs of discrete
invariance subgroups $C_n$ and $s=d([l_i])$ for compact versions of
$sl_{pd}(2)$ that is the case when $m\neq 0$ in (3.2). (Compact ($su_{pd}(2)$)
and non-compact ($su_{pd}(1,1)$) realzations of $sl_{pd}(2)$ algebras are
distinguished depending on whether dimensions $d([l_i])$ of the spaces
$L([l_i])$ are finite or infinite [12].) Eqs. (3.1), (3.4)-(3.6) yield
requisites for developing both exact and approximate Lie-algebraic methods to
solve physical tasks in models (3.2). We outline them following the papers
[12,13].

{\it Exact methods} are based on using Eqs. (3.1) and their resemblances with
defining relations for $sl(2)$. Thus, substituting Eq.(3.4) in the
Heisenberg equations for cluster dynamic variables $V_{\alpha}(t)$ related to
generators $V_{\alpha}$ one gets non-linear equations
\begin{equation}
i\frac{d V_0}{dt}=g V_+-g^*V_-,\; i\frac{d V_+}{dt}=-aV_+-g^*\psi^{p-1}(V_0),
\; i\frac{d V_-}{dt}=aV_- +g\psi^{p-1}(V_0)
\end{equation}
generalizing linear Bloch equations for $sl(2)$ and having in the cluster
mean-field approximation ($<|f(V_{\alpha})|>=f(<|V_{\alpha}|>$) quasiclassical
solutions expressed in terms of hyperelliptic (Abelian) functions [12].
However, such direct extensions of $sl(2)$-algebraic techniques [4] are
impossible for finding evolution ($U_{H}(t)$) and diagonalizing ($S$)
operators because relevant disentangling formulas for$\exp(\sum_i a_iV_i)$
and explicit expressions for matrix elements $\langle [l_i]; f|\exp(\sum_i
a_i V_i)|[l_i]; v\rangle$ are absent [12].

Nevertheless, substituting "pseudogroup" (cf. [9]) representations
\begin{equation}
U_{H}(t)=\sum_{f=-\infty}^{\infty} V_+^f \,u^{H}_f (V_0;t),\;
S=\sum_{f=-\infty}^{\infty} V_+^f\, S_f(V_0)
\end{equation}
%%%
for $U_{H}(t)$ and $S$ (with $V_+^{-k}\equiv V_-^k\left [\prod_{l=0}^{k-1}
\Psi^p(V_0-l)\right ]^{-1}$ due to Eq. (3.5)) or expansions
\begin{equation}
|E([l_i];f)\rangle = A_f\prod_{j}(V_+ -\Lambda^f_j(V_0))|[l_i]\rangle=
A_f\prod_{j}(V_+ -\kappa^f_j)|[l_i]\rangle=\sum_v Q_v(E_f) |[l_i]; v\rangle
\end{equation}
%%%%%%%%
for energy eigenstates $|E([l_i];f)\rangle$ in the diagonalizing scheme
$SHS=\tilde H(V_0)$ and the time-dependent Schroedinger equation
$i\hbar dU_{H}(t)/d t=HU(t)$ and using Eqs. (3.1), (3.5), one gets
finite-difference and differential-difference equations determining (together
with unitarity conditions) "coefficients" $S_f(Y_0)$, $u^{H}_f(V_0;t)$,
diagonal Hamiltonian forms $\tilde H(V_0)$, amplitudes $Q_v(E_f)$ and energy
spectra $\{E([l_i];f)\}$ [12,13]. (Note that two first equalities in (3.9)
realize, in fact, the algebraic Bethe ansatz [17] for wave functions
$|E([l_i];f)\rangle$ in terms of the $sl_{pd}(2)$ generators [13].) These
equations define new (non-classical) orthogonal functions in both
discrete and continious variables which are simultaneously related to
solutions of singular differential equations yielded by using two conjugate
differential realizations of generators $V_{\alpha}$ [12,13]: \,$V_+=z,\;
V_0=zd/dz+l_0,\; V_-= z^{-1} \Psi(zd/dz+l_0)$ and $V_-=d/dz,\; V_0=zd/dz+l_0,
\; V_+=\Psi(zd/dz+l_0)(d/dz)^{-1}$ generalizing the Bargmann or GCS
representations [2,4] for $sl(2)$ generators. However, at present, simple
analytical expressions for these special functions are absent in general
cases though some integral representations were found for them with the help
of a specific "dressing" of $sl(2)$-solutions of certain auxiliary exactly
solvable tasks [12].

{\it Approximate methods} developed in [12,13] are based on realizations of
generators $V_{\alpha}$ as special elements of extended enveloping algebras
${\cal U}_{\Psi}(sl(2))$ of the $sl(2)$ algebra via a generalized
Holstein-Primakoff map given on each subspace $L([l_i])$ as follows [12]
%%%%
\begin{equation}
V_0 = Y_0+l_0\pm J,\; V_+= Y_+ [\Phi (Y_0)]^{1/2},\;
\Phi (Y_0)=\frac{\Psi^p (V_0+1)}{\Psi^2 (Y_0+1)},\; V_-=(V_+)^+
\end{equation}
where $Y_{\alpha}$ and $\Psi^2 (Y_0)=(J\pm Y_0)(\pm J+1-Y_0)$ are generators
and the structure polynomials of the $su(2)/su(1,1)$ algebras, $\mp J$ are
lowest weights of their IRs realized on $L([l_i])$ (upper/lower signs
correspond to $su_{pd}(2)/su_{pd}(1,1)$ versions of $sl_{pd}(2)$).

Eqs. (3.10) enable us to re-write restrictions $H_{[l_i]}$ of Eqs. (3.4) on
$L([l_i])$ and basis vectors $|[l_i];v\rangle$ in terms of $Y_{\alpha}$:
%%%
\begin{equation}
H_{[l_i]}(\{Y_{\alpha}\})= aY_0+g Y_+\sqrt{\Phi(Y_0)}+ g^*\sqrt{\Phi(Y_0)}Y_-
+\tilde C,\; |[l_i];v\rangle={\cal N}(J,v)(Y_+)^v|[l_i]\rangle
\end{equation}
%%%%%
(with $\tilde C=C([l_i])+a(l_0\pm J)$) and to use the formalism [4] of the
$SL(2)$ GCS
\begin{equation}
|[l_i];v;\xi\rangle=S_Y|[l_i];v\rangle=\sum_{f\geq 0} S^Y_{f v}(\xi)
|[l_i];f\rangle, \; S_Y(\xi)\equiv\exp(\xi Y_+-\xi^* Y_-)
 \end{equation}
(with $S^Y_{f v}(\xi=re^{i\theta})$ being the $SL(2)$ Wigner $D$-function)
for examining models (3.4) in $SL(2)$-cluster (taking into account mode
correlations in (3.2)) quasiclassical approximations [13]. Specifically,
$SL(2)$ GCS determine quasiclassical energy functionals
%%%
\begin{equation}
\begin{array}{c}
{\cal H}^{cq}([l_i];v;\xi)=\langle[l_i];v;\xi|H_{[l_i]}|[l_i];v;\xi\rangle= \\
\tilde C+a(v\mp J) c(2r)-[g e^{-i\theta}+g^* e^{i\theta}]\sum_{f\geq 0}
|S_{f v}(\xi) S_{f+1 v}(\xi)|[\Psi^p (l_0+1+f)]^{1/2}
\end{array}
\end{equation}
%%%%%
(with $c(r)=\cos r/\cosh r\; \mbox{for}\; su(2)/su(1,1)$) or their mean-field
approximations
\begin{equation}
\begin{array}{c}
{\cal H}^{cmf}([l_i];v;\xi)=H_{[l_i]}(\langle[l_i];v;\xi|\{Y_i\}|[l_i];v;
\xi\rangle))= \\
\tilde C+a(v\mp J) c(2r)-[g e^{-i\theta}+g^* e^{i\theta}](J\mp v)s(2r)
[\Phi((\mp J+v)c(2r))]^{1/2}
\end{array}
\end{equation}
(with $c(r)=\cos r/\cosh r, s(r)=\sin r/\sinh r\;\mbox{for}\; su(2)/su(1,1)$)
which can be applied in standard calculation schemes [2,4,13].

For instance, inserting Eqs. (3.13) and (3.14) in the stationarity conditions
\begin{equation}
a)\;\frac{\partial {\cal H}([l_i];v;\xi)}{\partial \theta}=0,\qquad
b)\;\frac{\partial {\cal H}([l_i];v;\xi)}{\partial r}=0
\end{equation}
%%%%%%%
one finds approximate eigenfunctions $|E^{cq/cmf}([l_i];v)\rangle=
S_Y(\xi_0)^{\dagger}|[l_i];v \rangle$ and appropriate eigenenergies
$E^{cq}([l_i];v) ={\cal H}^{cq}([l_i];v;\xi_0),\, E^{cmf} ([l_i];v)=
{\cal H}^{cmf}([l_i];v;\xi_0)$ where $\xi_0=r_0g/|g|$ and values $r_0$
depend on $g, a, l_i$ and are determined by real solutions of algebraic
equations obtained from Eq. (3.15b) with $v=0$ [13].

The energy functionals (3.13), (3.14) may be also used for a quasiclassical
analysis of the $SL(2)$-cluster dynamics described by the classical
Hamiltonian equations (cf. [4])
%%%%%
\begin{equation}
\dot q =\frac{\partial {\cal H}}{\partial p}, \qquad \dot p =
-\frac{\partial {\cal H}}{\partial q}, \qquad {\cal H}=
\langle z(t);[l_i]|H|[l_i];z(t)\rangle
\end{equation}
%%%%%%%%
for the canonical parameters $p=\langle z(t);[l_i]|Y_0|[l_i];z(t)\rangle=\mp
J c(2r), q=\theta$ of the $SL(2)$ GCS $|[l_i];z(t)=r\exp (-i\theta)\rangle
=S_Y(\xi=-z(t))|[l_i]\rangle$ determining "principal" parts in the evolution
operators $U_{H}(t)= \exp(i\alpha (t) Y_0) S_Y(\xi (t))$ for initial states
$|\phi (0])\rangle=|[l_i]\rangle$[13]. Similarly, this approach yields
nonlinear quasiclassical Bloch-type equations
%%%%%
\begin{equation}
\dot {\bf y}=\frac{1}{2} {\bf \bigtriangledown} {\cal H}\times
{\bf \bigtriangledown}{\cal C}, \;{\bf y}=(y_1,y_2,y_0),\,{\cal C}=\pm y_0^2+
y_1^2+y_2^2 ,\, {\bf \bigtriangledown} =(\partial/\partial y_1,
\partial/\partial y_2,\partial/\partial y_0)
\end{equation}
which are equivalent in the mean-field approximation (3.14) to those obtained
from Eqs. (3.7) ($y_i=\langle[l_i];z|\{Y_i\}|[l_i];z\rangle,
{\bf A}\times{\bf B}$ is the vector product symbol).

So, Eqs. (3.11)-(3.17) yield $SL(2)$-cluster quasiclassical solutions of
spectral and evolution problems which take into account quantum correlations
of interacting subsystems though they do not describe quantum dynamics
exactly unlike the case of $sl(2)$-linear Hamiltonians [13].

\section{\hspace{-4mm}.\hspace{2mm}POLYNOMIAL OSCILLATOR
LIE ALGEBRAS in ANALYSIS of COMPOSITE FIELD MODELS}

\hspace*{0.8cm}
%Applications of algebras
%$osc_{pd}^{Y/X}(m;n)$ are discussed to construct
%field-theoretic composite models with internal symmetries where
%collective variables $Y^{+}/Y$ ($X^{+}/X$) are considered as
%creation/destruction operators of specific coherent $G_{i}$-invariant
%clusters. Then, from $Y^{+}/Y$ ($X^{+}/X$) we construct creation/destruction
%operators of particle-like excitations via a generalized Holstein-Primakoff
%map (3.12).
Another natural area of applications of algebras $g_{pd}$ is an algebraic
analysis [5,18] of composite field models with internal (gauge) symmetries [2]
which generalizes basic ideas of the paraquantization [3]. The simplest
example of such analysis (but without introducing algebras $g_{pd}$) was
given in [18] by using models (3.2) with $m=0=\omega_0$ to describe resonance
states in particle physics; later it was generalized on multimode cases and
applied to study multiphoton processes in quantum optics [5,10].

It was shown that operators $V^+=(a^+_1)^n$ describe $C_{n}$-invariant
n-particle kinematic clusters which display unusual (para)statistics and
correspond to generalized asymptotically free fields realized on the Fock
spaces $L_F$. Specifically, $V^+$-clusters satisfy non-canonical CRs (3.1)
and multi-linear relations [5]:
\begin{equation}
ad^{n+1}_{V}V^+=0,\quad ad_{V}V^+\equiv [V,V^+],\quad V=(a_1)^n=(V^+)^+
\end{equation}
generalizing (for $n\geq 3$) trilinear parastatistical Green's relations [3].
Furthermore, the subspaces $L([l_0=\kappa/n])$ in (2.2) describe coherent
mixtures of constant numbers $\kappa$ of uncoupled particles $a_1^+$ and of
varying in time numbers $N_V$ of $V$-clusters. However, operators $N_V$ have
not standard (for (papra)fields) bilinear in $V, V^+$ forms (cf. [2,3]) but
they can be expressed as nonlinear functions in the bilineals $V^+V, VV^+$
[5]:
\begin{equation}
N_V = (E_{11}- C(R_0))/n, \qquad E_{11}=a_1^+a_1=nV_0=\varphi (V^+V),
\quad C([l_0])=nl_0=\kappa
\end{equation}
as it follows from Eqs. (3.1), (3.5). Therefore, at best the quantities
$V^+$ can be set in correspondence only to parafield quanta [3] (when $n=2$)
rather than to any asymptotically free particles [5]. Nevertheless, one can
construct operators $W^+=W^+(\{V_i\}), W=(W^+)^+$ obeying canonical CRs
$[W, W^+]=1$ and having the standard number operators $N_W=W^+W(=N_V)$.
Specifically, in [5,18,12] two equivalent forms were found for $W^+$:
%%%%%
\begin{equation}
W^+ = V^+\sum_{r\geq 0} c_r (V^+)^r(V)^r= \,V^+\left [\frac{V_0-R_0+1}
{(E_{11}+n)^{(n)}}\right ]^{1/2}
\end{equation}
where the second form is, in fact, a modification of the (inverted) map
(3.10) [12].

The analysis above was generalized in [5] by means of: 1) extending on
the case of $"m"$ modes using $C_n$-invariant Hamiltonians
\begin{equation}
H = C + \sum_{i=1}^m \omega_i a_i^+ a_i + \sum_{1\leq i_1\dots\leq m}
[g_{i_1\dots}V^+_{i_1\dots} + g^*_{i_1\dots} V_{i_1\dots}], \,V^+_{i_1\dots}=
a_{i_1}^+\dots a_{i_n}^+ ;
\end{equation}
2) involving both boson ($a_i$) and fermion ($b_i$) variables; 3) considering
Hamiltonians with non-Abelian invariance groups $G_i=SU(n)$ (obtained from
(4.4) by the substitutions: $a_i^+ a_i\rightarrow({\bf a_i^+}\cdot{\bf a_i})
\equiv\sum_{j=1}^n a_{ji}^+ a_{ji},\; V^+_{i_1\dots}\rightarrow
X^+_{i_1\dots}\equiv\sum_{(j_k)}\epsilon_{j_1 \dots j_n} a_{j_1 i_1}^+
\dots  a_{j_n i_n}^+$). These procedures determine generalized dual pairs
($C_{n}, osc_{pd}^{V}(m;n)$) and ($SU(n), osc_{pd}^{X}(m;n)$) where
polynomial oscillator (super)algebras
$osc^{V}(m;n)$ and $osc^{X}(m;1^n)$ are extensions of the unitary algebras
$u(m)=Span\{ E_{ij}=a_i^+ a_j\}\, \mbox{and}\, Span\{ E_{ij}=
({\bf a_i^+}\cdot{\bf a_j})\}$ by their symmetric ($V^+_{\dots}, V_{\dots}$)
and skew-symmetric ($X^+_{\dots}, X_{\dots}$) tensor operators.

The operators $X^+_{\dots},X_{\dots}$ and $V^+_{\dots},V_{\dots}$ satisfy
non-canonical CRs with right sides depending on $E_{ij}$ (and on the $SU(n)$
Casimir operators for $osc^{X}(m;1^n)$) and obey (due to the invariant
theory) certain extra "bootstrap" relations (of the type $V_{1\dots1}
V_{2\dots2}=V_{21\dots1}V_{12\dots2}$) [5] which are similar to those
occuring in quantum field theories with constraints [2,3].
All this entails unusual statistical and other properties of $G_i$-invariant
clusters associated with $X^+_{\dots}, V^+_{\dots}$ and complicates
extensions of the $1$-mode analysis above. Specifically, the task of
obtaining $m$-mode generalizations of the map (4.3),
%%%%%
\begin{equation}
W^+_a = \sum_{i_1\dots i_n} (V^+_{i_1\dots i_n}/X^+_{i_1\dots i_n} )
f^a_{i_1\dots i_n} (\{ E_{ij}\}),\qquad [W_a, W^+_b]=\delta_{a b},\; W_a=
(W^+_a)^+
\end{equation}
(with coefficients $f^a_{\dots}(...)$ determined from finite-difference
equations due to $[W_a, W^+_b]=\delta_{a b}$) is, in general, sufficiently
difficult because of relations abovementioned between $V/X$-clusters.

\section{\hspace{-4mm}.\hspace{2mm}CONCLUSION}

\hspace*{0.8cm} In conclusion we briefly point out some of prospects of
developing results obtained. For instance, formal constructions of Sections 2,
4 may be generalized by involving into consideration q-deformed oscillators
and other invariance groups $G_i$.
It is also of interest to examine infinite-dimensional algebras $\hat{g}_{pd}$
related to $g_{pd}$ along lines of standard studies [7,8] and to construct
non-Fock realizations of the $g_{pd}$ IRs (cf. [11]) extending arbitrary
$h$-modules with the help of coset generators.

It is of very importance to develop exact methods of Section 3 since they
outline ways of generalizing standard group-theoretical technigues for
solving both spectral and evolution tasks with dynamic symmetry algebras
$g_{pd}$. Specifically, one may use techniques of $q$-deformed algebras and
$q$-special functions [6] (due to interrelations between $sl_{pd}(2)$ and
$q$-deformed algebras) as well as pseudogroup and braided geometry concepts
[9,10] (due to Eqs. (3.8),(3.9)) for determining new classes of special
functions related to $g_{pd}$. At the same time quasiclassical approximations
obtained may be considered as specific asymptotics of exact solutions that
opens a possibility to use the techniques of asymptotic expansions [19] for
finding latters. On the other hand, solutions of the nonlinear Eqs. (3.7)
enable to determine operator analogs of Abelian functions which are,
probably, related to quantization problems on algebraic varieties [12].

Results of Section 4 provide an effective analysis of composite models with
internal $G_i$-symmetries at algebraic and quasi-particle levels, when
obtaining explicit expressions for $f_{\dots}(\dots)$ in Eqs. (4.5) and 
examining the limit "$m \rightarrow \infty$". Furthermore, involving into 
consideration spatiotemporal symmetries, one can construct appropriate
"physical" composite fields and relevant Hamiltonians (Lagrangians) for
them in terms of "quanta" $W_a$ (cf. [2,3,5,10]).

The author thanks A. Odzijewicz for interest in the work, useful
discussions and remarks and V.N. Tolstoy for stimulating discussions.
A partial support of the Russian Foundation for Basic Research (under grant
No 96-02 18746a) is acknowledged.

\end{document}